\documentclass[conference, a4paper]{IEEEtran}	
\IEEEoverridecommandlockouts

\usepackage[cmex10]{amsmath}
\usepackage{amssymb}
\usepackage{epsfig}
\usepackage{subfigure}
\usepackage{cite}
\usepackage{graphicx}
\usepackage{color}
\usepackage{bm}
\usepackage{booktabs}
\usepackage{gensymb}
\usepackage{mathrsfs}
\usepackage{xfrac}
\usepackage{algorithm}
\usepackage{algorithmic}
\usepackage{fancyhdr}
\usepackage{setspace} 
\newlength{\figwidth}
\makeatletter
\renewcommand\normalsize{%
   \@setfontsize\normalsize\@xpt\@xiipt
   \abovedisplayskip 5\p@ \@plus2\p@ \@minus3\p@
   \abovedisplayshortskip \z@ \@plus3\p@
   \belowdisplayshortskip 3\p@ \@plus3\p@ \@minus3\p@
   \belowdisplayskip \abovedisplayskip
   \let\@listi\@listI}
\makeatother

\usepackage{geometry}
\begin{document}
\title{PASS-Enhanced MEC: Joint Optimization of Task Offloading and Uplink PASS Beamforming}
\vspace{-2cm}
\author{\IEEEauthorblockN \normalsize {Zhaoming~Hu\IEEEauthorrefmark{1}\IEEEauthorrefmark{2}, Ruikang~Zhong\IEEEauthorrefmark{3}, Xidong~Mu\IEEEauthorrefmark{4}, Dengao~Li\IEEEauthorrefmark{1}\IEEEauthorrefmark{2}, Yuanwei~Liu,~\IEEEauthorrefmark{5}}
\IEEEauthorblockA{\IEEEauthorrefmark{1} \footnotesize College of Computer Science and Technology (College of  Data Science), Taiyuan University of Technology, Taiyuan 030024, P.R. China}
\IEEEauthorblockA{\IEEEauthorrefmark{2} \footnotesize The Key Laboratory of Data Governance and Intelligent Decision-Making of Shanxi Province, Taiyuan 030024, P.R. China}
\IEEEauthorblockA{\IEEEauthorrefmark{3} \footnotesize School of Information and Communication Engineering, Xi'an Jiaotong University, Xi'an 710049, P.R. China}
\IEEEauthorblockA{\IEEEauthorrefmark{4} \footnotesize Centre for Wireless Innovation (CWI), Queen's University Belfast, Belfast, BT3 9DT, U.K.}
\IEEEauthorblockA{\IEEEauthorrefmark{5} \footnotesize Department of Electrical and Electronic Engineering, The University of Hong Kong, Hong Kong}}
\vspace{-5cm}
\maketitle

\begin{abstract}
     A pinching-antenna system (PASS)-enhanced mobile edge computing (MEC) architecture is investigated to improve the task offloading efficiency and latency performance in dynamic wireless environments. By leveraging dielectric waveguides and flexibly adjustable pinching antennas, PASS establishes short-distance line-of-sight (LoS) links while effectively mitigating the significant path loss and potential signal blockage, making it a promising solution for high-frequency MEC systems. We formulate a network latency minimization problem to joint optimize uplink PASS beamforming and task offloading. The resulting problem is modeled as a Markov decision process (MDP) and solved via the deep reinforcement learning (DRL) method. To address the instability introduced by the $\max$ operator in the objective function, we propose a load balancing-aware proximal policy optimization (LBPPO) algorithm. LBPPO incorporates both node-level and waveguide-level load balancing information into the policy design, maintaining computational and transmission delay equilibrium, respectively. Simulation results demonstrate that the proposed PASS-enhanced MEC with adaptive uplink PASS beamforming exhibit stronger convergence capability than fixed-PA baselines and conventional MIMO-assisted MEC, especially in scenarios with a large number of UEs or high transmit power.
\end{abstract}

\vspace{-3mm}
\section{Introduction} \label{Introduction}

With the rapid evolution toward beyond fifth-generation (B5G) and sixth-generation (6G) wireless networks, emerging applications such as immersive metaverse, intelligent transportation, and industrial automation impose stringent requirements on both communication and computation performance. Massive numbers of heterogeneous user equipments (UEs) continuously generate intensive workloads that demand high data rates, ultra-low latency, and reliable service guarantees. Traditional cloud computing paradigm, although powerful in computational capability, is limited by long-distance backhaul transmission, which introduces excessive delays and is inadequate for delay-sensitive and computation-intensive applications. To address this issue, mobile edge computing (MEC) has been proposed as a promising paradigm that brings computing resources closer to UEs by deploying servers at the network edge, such as access points (APs) and base stations (BSs)~\cite{Gu2024COMST}. By offloading computation tasks to nearby MEC servers, mobile terminals can significantly reduce task execution latency and alleviate their energy consumption constraints, thereby improving the overall quality of service (QoS).


The performance of MEC is critically dependent on wireless link quality, which is susceptible to user mobility, blockages, and fading, making reliable connectivity a key challenge. Advanced antenna technologies are fundamental to enhancing link reliability and spectral efficiency. While conventional approaches like massive multiple-input multiple-output (MIMO)~\cite{Liu2021COMST} use spatial diversity to improve capacity, and emerging technologies like reconfigurable intelligent surfaces (RISs) dynamically shape the propagation environment, they face limitations in dynamic MEC scenarios. Issues such as hardware complexity, limited reconfiguration agility, and poor adaptability to mobile users motivate the development of more flexible and efficient wireless solutions.

As a groundbreaking architecture recently prototyped by NTT DOCOMO~\cite{DOCOMO}, the pinching-antenna system (PASS) offers a hardware-reconfigurable solution. It overcomes the key limitations of conventional multi-antenna systems, namely free-space path loss and line-of-sight (LoS) blockage. Unlike traditional approaches, PASS employs low-loss dielectric waveguides as its primary signal transport medium~\cite{Liu2025COM}. Small dielectric particles, referred to as pinching antennas~(PAs), can be dynamically attached or detached at arbitrary positions along the waveguide. This ``pinching'' mechanism enables flexible signal radiation into free space, effectively forming near-wired LoS links when PAs are placed in close proximity to users. Such a design not only drastically reduces propagation loss but also allows real-time adjustment of the number and layout of antennas without complex radio frequency (RF) circuitry. Recent studies have begun to explore the communication potential of PASS. \cite{Ren2025arxiv}~evaluated the performance of PASS with different multiple access technologies, while \cite{Zhu2025COM} investigated a secure wireless communication framework. These works highlight the advantages of PASS in supporting scalable, low-cost, and high-quality connectivity, making it a highly suitable candidate for enhancing MEC systems in dynamic wireless environments.


Motivated by the above observations, this paper investigates a PASS-enhanced MEC system, where a network latency minimization problem is formulated to joint optimize uplink PASS beamforming and task offloading. As a long-term stochastic optimization, the problem can be modeled as a Markov decision process (MDP) and effectively addressed using deep reinforcement learning (DRL), which is well suited for sequential decision-making in dynamic environments. Nevertheless, the presence of $\max$ operator in the objective function often leads to instability and slow convergence. To overcome this issue, we propose a load balancing-aware proximal policy optimization (LBPPO) algorithm. By incorporating both node-level and waveguide-level load balancing information into the policy design, LBPPO enhances training stability and convergence, and improving latency performance in PASS-assisted MEC systems.

\section{System Model and Problem Formulation} \label{system_model}

\begin{figure}[h]
\centering
\includegraphics[scale=0.5]{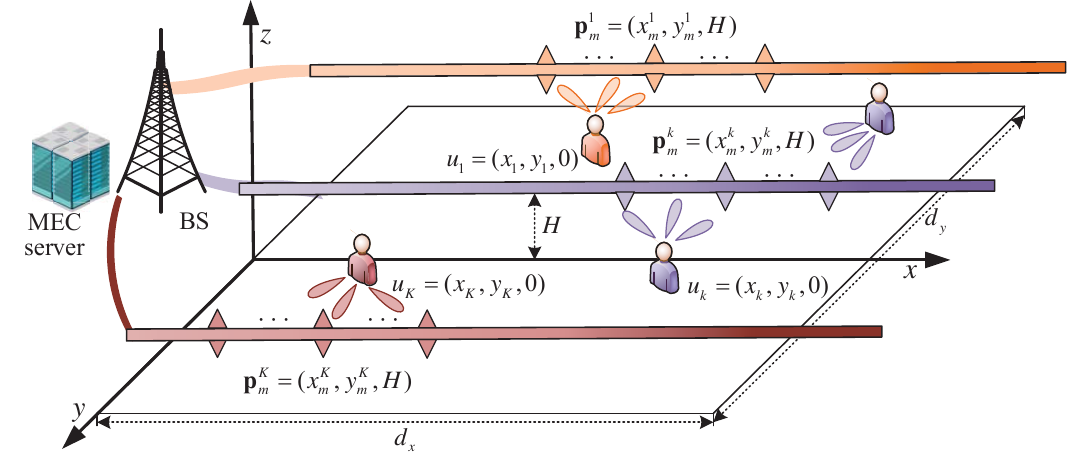}
\caption{System model of PASS-enabled MEC system.}
\label{fig:system model}
\vspace{-0.4cm}
\end{figure}

As illustrated in Fig.~\ref{fig:system model}, we consider a PASS-enhanced wireless MEC system. The system consists of a BS equipped with $N$ waveguides and $K$ UEs with $L$ antennas. In addition, there are $M$ PAs on each waveguide. Considering a three-dimensional (3D) Cartesian coordinate system, we assume that the users are located on the $x - y$ plane with a squared area of $d_x \times d_y$, and the system operates over an extended temporal horizon comprising $T$ discrete time slots. The $\mathbf{p}_{k,\tau} = (x_{k,\tau},y_{k,\tau},0)$ denotes the position of the $k$-th UE at time slot $t$, where $\forall k \in \mathcal{K}=\left\{1,2,...,K\right\}$, and $\forall \tau \in \mathcal{T}=\left\{1,2,...,T\right\}$. Specifically, $x_{k,\tau}$ follows a uniform distribution in $\left[0,d_x\right]$ and $y_{k,\tau}$ follows a uniform distribution in $\left[-\frac{d_y}{2},\frac{d_y}{2}\right]$. Without the loss of generality, it is assumed that the waveguides with the length of $d_x$ is deployed at the height of $H$ and in parallel to the $x$-axis. The location of the $m$-th PA on the $n$-th waveguide at time slot $\tau$ is denoted as $\mathbf{p}_{m,n,\tau} = (x_{m,n,\tau},y_{m,n,\tau},H)$, and all $m$ PAs on the same waveguide $n$ share the same $y$-coordinate, i.e., $y_{1,n,\tau}=y_{2,n,\tau}=...=y_{M,n,\tau}$ for all $n \in \mathcal{N}$ and $\tau \in \mathcal{T}$.



In the proposed system, UEs are equipped with wireless communication circuits and computing processors that have limited computational capabilities, and they handle computation tasks involving substantial amounts of input data (measured in bits). These tasks can be processed locally by the UEs or offloaded to the BS, which is equipped with an MEC server. The PASS, which can transmit an incident signal, is deployed to assist the UEs' task offloading. Given the negligible bandwidth consumption of downlink result delivery in MEC systems, our subsequent analysis will focus on the uplink PASS scenario.

\subsection{Signal Model}
In consideration of large-scale channel characteristics, the analysis is confined to LoS channels to explore fundamental performance limits. Within this framework, the small-scale fading model is simplified to focus on beamforming dynamics. The channel gain is normalized to exclude large-scale path loss, while phase variations induced by relative UE movement are explicitly modeled through propagation distance differences across the antenna array. Specifically, consider a uniform linear array (ULA) of $L$ antennas at the UE, aligned along the $y$-axis with its centroid at $\mathbf{p}_{k,\tau}$. The location of the $l$-th antenna is then given by 
\begin{align}
\mathbf{p}_{k,l,\tau} = \left(x_{k,\tau},\ y_{k,\tau} + \delta_l,\ 0\right),\ \delta_l = \left(l - \frac{L+1}{2}\right)\Delta_l,
\end{align}
where $\Delta_l$ denotes the UE's antenna spacing. Under these conditions, the channel matrix $\mathbf{H}_{k,n,\tau} \in \mathbb{C}^{M \times L}$ between the PA on the $n$-th waveguide and the $k$-th UE can be expressed as
\begin{align}
&\mathbf{H}_{k,n,\tau} = \\ \nonumber &\left[
\begin{array}{ccc}
\frac{\sqrt{\eta}e^{-j\frac{2\pi}{\lambda}|\mathbf{p}_{k,1,\tau}-\mathbf{p}_{1,n,\tau}|}}{|\mathbf{p}_{k,1,\tau}-\mathbf{p}_{1,n,\tau}|} & \cdots & \frac{\sqrt{\eta}e^{-j\frac{2\pi}{\lambda}|\mathbf{p}_{k,L,\tau}-\mathbf{p}_{1,n,\tau}|}}{|\mathbf{p}_{k,L,\tau}-\mathbf{p}_{1,n,\tau}|} \\
\vdots & \ddots & \vdots \\
\frac{\sqrt{\eta}e^{-j\frac{2\pi}{\lambda}|\mathbf{p}_{k,1,\tau}-\mathbf{p}_{M,n,\tau}|}}{|\mathbf{p}_{k,1,\tau}-\mathbf{p}_{M,n,\tau}|} & \cdots & \frac{\sqrt{\eta}e^{-j\frac{2\pi}{\lambda}|\mathbf{p}_{k,L,\tau}-\mathbf{p}_{M,n,\tau}|}}{|\mathbf{p}_{k,L,\tau}-\mathbf{p}_{M,n,\tau}|}
\end{array}\right],
\end{align}
where $\sqrt{\eta}=\frac{c}{4\pi f_c}$ with $c$ and $f_c$ represent the speed of light and the carrier frequency, respectively. $\lambda$ denotes the wavelength in free space. $|\mathbf{p}_{k,\tau}-\mathbf{p}_{m,n,\tau}|$ represents the distance between the $k$-th user and $m$-th PA on the $n$-th waveguide at time slot $\tau$, which can be modeled as
\begin{align}
|\mathbf{p}_{k,l,\tau}-& \mathbf{p}_{m,n,\tau}| \\ \nonumber = &\sqrt{(x_{k,\tau}-x_{m,n,\tau})^2 + (y_{k,\tau} + \delta_l - y_{m,n,\tau})^2 + H^2},
\end{align}

For the $m$-th PA on the $n$-th waveguide, as all PA components share a unified transmission medium, the radiated signals from each antenna element inherently manifest as phase-adjusted duplicates of the BS's excitation signal propagating from the $n$-th waveguide's feed point, which can be formulated as
\begin{align}\label{h_n_single}
\mathbf{h}_{n,\tau} = \left[e^{-j\frac{2\pi}{\lambda_w}|\mathbf{p}_{0,n,\tau}-\mathbf{p}_{1,n,\tau}|},\ldots, e^{-j\frac{2\pi}{\lambda_w}|\mathbf{p}_{0,n,\tau}-\mathbf{p}_{M,n,\tau}|} \right]^T,
\end{align}
where $\mathbf{p}_{0,n,\tau}$ represents the coordinates of the feed point of the $n$-th waveguide, and $\lambda_w = \lambda / n_e$ represents the guided wavelength with $n_e$ corresponds to the dielectric waveguide's effective refractive index.

In the considered uplink system, each multi-antenna UE leverages beamforming capabilities to selectively associate with a set of PAs on one waveguide, thereby optimizing task offloading efficiency. By exploiting spatial diversity through coordinated beamforming, UEs dynamically align their antenna arrays to specific PAs. This alignment enhances channel conditions via coherent signal combining and reduces inter-user interference. We refer to this strategy as \textbf{\emph{uplink PASS beamforming}}. Such association mechanisms are particularly advantageous in MEC systems, as they enable reliable low-latency transmission of compute-intensive tasks by maximizing the effective signal-to-interference noise ratio (SINR) at the receiver. Furthermore, adaptive waveguide-UE association facilitates load balancing of PASS-assisted wireless transmissions, ensuring efficient utilization of the MEC server's (including UE and BS) computational resources while mitigating latency for data offloading transmission. Let $\mathbf{w}_{k,\tau} \in \mathbb{C}^{L \times 1}$ denote the UE's transmit beamforming vector at the $k$-th UE and time slot $\tau$, where $||\mathbf{w}_{k,\tau}||^2 \leq P_{k}^{\text{max}}$. The association between the $k$-th UE and the $n$-th waveguide at time slot $\tau$ is governed by a binary variable $\alpha_{k,n,\tau} \in \left\{0,1\right\}$, where $\alpha_{k,n,\tau}=1$ indicates active association between UE $k$ and waveguide $n$. Each user can only be associated with one waveguide, i.e., $\sum_{n \in \mathcal{N}} \alpha_{k,n,\tau} = 1$. This user-to-waveguide allocation follows the waveguide division multiple access (WDMA) paradigm introduced for PASS in~\cite{Zhao2025arxiv}. The BS decodes each UE's signal based solely on the reception from its associated waveguide, forgoing any joint processing across the $N$ received signal streams. The composite received signal at the BS from UE $k$ is then expressed as
\begin{align}\label{eq:multi_antenna_model}
y_{k,\tau} = \sum_{n=1}^N \alpha_{k,n,\tau} \left(\underbrace{\mathbf{h}_{n,\tau}^H \mathbf{H}_{k,n,\tau}}_{\substack{\text{Pinching} \\ \text{beamforming}}} \underbrace{\mathbf{w}_{k,\tau}}_{\substack{\text{Transmit} \\ \text{beamforming}}} s_{k,\tau} + n_{k,\tau} \right),
\end{align}
where $s_{k,\tau}$ represents the symbol of the $k$-th UE transmitted signal and $n_{k,\tau}$ represents the aggregated additive white Gaussian noise (AWGN) on the RF chain of the waveguide associated with $k$-th UE. Let the noise from the $M$ PAs be independent, each with a power of $\sigma^2$. The resulting aggregated noise power at the feed point of the waveguide associated with $k$-th UE is thus $M\sigma^2$, where $n_{k, \tau} \sim \mathcal{CN}(0, M\sigma^2)$. 
When decoding the task-offloading data of UE $k$, the effective received SINR is determined by
\begin{equation}\label{eq:sinr}
    \gamma_{k,\tau} = \dfrac{
        \left| \sum_{n=1}^N \alpha_{k,n,\tau} \mathbf{h}_{n,\tau}^{H} \, \mathbf{H}_{k,n,\tau} \, \mathbf{w}_{k,\tau} \right|^2
    }{
    \sum_{j \neq k} \left| \sum_{n=1}^N \alpha_{k,n,\tau} \mathbf{h}_{n,\tau}^T \mathbf{H}_{j,n,\tau} \mathbf{w}_{j,\tau} \right|^2  + M \sigma^2,
}
\end{equation}
where $M\sigma^2$ represents the aggregated noise power.

Eventually, the achievable communication rate of UE $k$ is given by
\begin{align}
R_{k,\tau} = B_{k,\tau} {\rm log}_2 (1 + \gamma_{k,\tau}),
\end{align}
where $B_{k,\tau}$ represents bandwidth allocated to the $k$-th UE at time slot $\tau$. We assume that all waveguides share the same frequency band and the total channel bandwidth is allocated evenly to all UEs. Thus, the bandwidth allocated to the UE $k$ is expressed as $B_{k,\tau} = \frac{B_{\text{total}}}{K}$, where $B_{\text{total}}$ denotes the total bandwidth of the system.
\vspace{-0.2cm}
\subsection{Task Offloading}
The proposed system employs a partial computation offloading scheme, where computational tasks generated by UEs are divisible and dynamically distributed between local execution and BS offloading through PASS transmission. Specifically, the scheme enables: 1) local processing of a task subset at the UE; 2) wireless transmission of remaining tasks to the BS via PASS for remote computation. Furthermore, the system operates over an extended temporal horizon, with the UE initiating service requests at each slot. However, constrained by the finite computational capabilities of both the UE and BS, processing queues inevitably accumulate within their respective server buffers. This queuing dynamic necessitates careful resource allocation to maintain system stability throughout the operational period. For any task generated by UE $k$ at time slot $\tau$, the local computation latency $t_{k,\tau}^l$ consists of the local queuing latency of the task waiting in the queue $t_{k,\tau}^{l,q}$ and local computation latency $t_{k,\tau}^{l,c}$, which is denoted as
\begin{align}
t_{k,\tau}^l = t_{k,\tau}^{l,q} + t_{k,\tau}^{l,c}.
\end{align}

For the $k$-th UE, $f_k$ represents its computing capability. Meanwhile, $L_{k,\tau}$ denotes the total size (in bits) of the computation task generated by UE $k$ at time slot $\tau$. Let $\beta_{k,\tau}$ denote the offloading ratios of UE $k$. We assume that the MEC servers at UEs and BS use first-input first-output (FIFO) to process tasks in the queue, but they have inherent limitations on their service capacity. Therefore, the offloading decision in time slot $\tau-1$ will affect the state of the queue at the UE and BS at time slot $\tau$. The MDP will be employed to construct the queuing model. The local queuing length and latency of the task generated by UE $k$ at time slot $\tau$ are expressed as
\begin{align}
Q_{k,\tau} =
\begin{cases}
0, & \tau = 1 \\
\max\left( Q_{k,\tau-1} + \lambda_{k,\tau-1} - \dfrac{f_k \cdot \dot{\tau}}{\rho}, 0 \right), & \tau > 1
\end{cases}
\end{align}
\begin{align}
t_{k,\tau}^{l,q} = \frac{Q_{k,\tau} \cdot \rho}{f_k},
\end{align}
where $Q_{k,\tau}$ and $\dot{\tau}$ represent the size of task data in UE $k$'s local queue and the duration of a time slot (in seconds) in the physical world, respectively. $\lambda_{k,\tau-1}=(1-\beta_{k,\tau-1})L_{k,\tau-1}$ denotes the size of the locally processed task generated by UE $k$ at time slot $\tau-1$. $f_k$ represents the operating frequency of the central processing unit (CPU) of UE $k$, measured in Hertz. $\rho$ is the computation density, i.e., the number of CPU cycles required to compute one bit of task data.
In addition, the local computation latency of the task generated by UE $k$ at time slot $\tau$ is expressed as
\begin{align}
t_{k,\tau}^{l,c} = \frac{(1-\beta_{k,\tau})L_{k,\tau}\rho}{f_k}.
\end{align}


For any task generated by UE $k$ in time slot $\tau$, the offloading latency consists of the offloading transmission latency $t_{k,\tau}^{o,\tau}$ of the PASS system, the offloading queuing latency $t_{k,\tau}^{o,q}$ and the offloading computational latency $t_{k,\tau}^{o,c}$ at the BS, which can be denoted by
\begin{align}\label{t_o}
t_{k,\tau}^o = t_{k,\tau}^{o,tr} + t_{k,\tau}^{o,q} + t_{k,\tau}^{o,c}.
\end{align}

In (\ref{t_o}), the queueing length, queuing latency, and computation latency of the task generated by UE $k$ at time slot $\tau$ are expressed as
\begin{align}
Q_{b,\tau} =
\begin{cases}
0, & \tau = 1, \\
\max\left( Q_{b,\tau-1} + \lambda_{b,\tau-1} - \dfrac{f_b \cdot \dot{\tau}}{\rho}, 0 \right), & \tau > 1,
\end{cases}
\end{align}
\begin{align}
t_{k,\tau}^{o,q} = \frac{Q_{b,\tau} \cdot \rho}{f_b},
\end{align}
\begin{align}
t_{k,\tau}^{o,c} = \frac{\beta_{k,\tau}L_{k,\tau}\rho}{f_b},
\end{align}
where $Q_{b,\tau}$ and $f_b$ represent the size of task data in BS's queue at time slot $\tau$ and computing capability of BS, respectively. $\lambda_{b,\tau-1}=\sum_{k=1}^K \beta_{k,\tau-1}L_{k,\tau-1}$ denotes the size of the offloading tasks in the BS at time slot $\tau$. $\kappa_b$ is the effective capacitance coefficient of the BS.
Furthermore, for the offloading transmission latency, the additional latency of the system due to PA movement between time slots needs to be considered. Therefore, the overall offloading transmission latency of the task generated by UE $k$ at time slot $\tau$ is formulated as
\begin{align}
t_{k,\tau}^{o,tr} = \frac{\beta_{k,\tau} L_{k,\tau}}{R_{k,\tau}} + t^{\text{mov}}_\tau,
\end{align}
\begin{align}
t^{\text{mov}}_\tau = \omega(x_{m,n,\tau}-x_{m,n,\tau-1}),
\end{align}
where $t^{\text{mov}}_\tau$ and $\omega$ represent latency and latency consumption factor per unit distance of PA movement, respectively. In addition, the movement of PA will affect the offloading latency of each user's computing request.

Given the typically small-sized computational results, the corresponding feedback latency is considered negligible in this configuration. As local processing and data offloading operate concurrently, the total task execution latency for UE $k$ is determined by the slower of the two parallel processes, which is formulated as
\begin{align}
t_{k,\tau} = \max(t_{k,\tau}^l, t_{k,\tau}^o ).
\end{align}


\subsection{Problem Formulation}

Our objective is to optimize the UE's acitve beamforming vector $\mathbf{w}$, the positions of PAs $\mathbf{P}$, the UE and waveguide association vector $\boldsymbol{\alpha}$, and offloading decision vector $\boldsymbol{\beta}$ with the aim of minimizing the overall network latency. In pursuit of this goal, we formulate the optimization problem as
\begin{subequations}
\begin{align}\label{OPP1}
\textbf{P}:  \min_{\mathbf{w}, \mathbf{P}, \boldsymbol{\alpha}, \boldsymbol{\beta}} & \sum_{\tau=1}^{T} \sum_{k=1}^{K} t_{k,\tau},\\
\textrm{s.t.} \; \ \
& |x_{m,n,\tau} - x_{m^\prime,n,\tau}| \leq \Delta_l, \forall m, m^\prime, \forall n, \forall \tau,\label{OPP1A}\\
& x_{m,n,\tau} \in [0, d_x], \forall m, \forall n, \forall \tau, \label{OPP1B}\\
& y_{1,n,\tau}=y_{2,n,\tau}=...=y_{M,n,\tau}, \forall n, \forall \tau,\label{OPP1C}\\
& t_{k,\tau} \leq t_{\text{QoS}}, \forall k, \forall \tau, \label{OPP1D}\\
& ||\mathbf{w}_{k,\tau}||^2 \leq P_{k}^{\text{max}}, \forall \tau\label{OPP1E}\\
& \sum_{n \in \mathcal{N}} \alpha_{k,n,\tau} = 1, \forall k, \forall \tau, \label{OPP1F}\\
& \alpha_{k,n,\tau}, \beta_{k,\tau} \in \{0,1\} \forall k, \forall n, \forall \tau. \label{OPP1G}
\end{align}
\end{subequations}

Constraint \eqref{OPP1A} ensures that the distance between adjacent PAs is great than or equal to $\Delta$ to avoid coupling effects. Based on the inherent physical connectivity of the PA to the waveguide, \eqref{OPP1B} and \eqref{OPP1C} constrain the position of the PAs. Constraint \eqref{OPP1D} is a QoS constraint specifically the maximum waiting time of UE. Constraint \eqref{OPP1E} limits the transmit power of the UE beamforming. Additionally, \eqref{OPP1F} pertain to user association constraints. Constraint \eqref{OPP1G} requires that all the boolean variables can only be assigned values of 0 or 1.

\section{Load Balancing-Aware PPO Optimization Design for PASS-enhanced MEC System} \label{LBPPO}


The proposed PASS-enhanced MEC system operates over a long-term horizon, where offloading decisions in each time slot affect the queue states of both UEs and the BS. Considering these temporal dependencies and the non-negligible overhead from PA movements between slots, the optimization problem \textbf{P} is formulated as a MDP. By leveraging deep neural networks as function approximators, DRL is employed to handle the high-dimensional state and action spaces while capturing long-term temporal correlations inherent in the joint optimization of task offloading and uplink PASS beamforming.


However, in latency-oriented optimization of MEC, objective functions involve max operators across heterogeneous nodes, leading to non-smooth gradients that hinder convergence in conventional methods. Proximal policy optimization (PPO), a policy-gradient-based DRL algorithm, overcomes this issue with a clipped surrogate objective that limits policy updates to a trust region. This approach enhances training stability and convergence on non-smooth objectives while balancing exploration and exploitation. Such characteristics make PPO well-suited for non-convex constrained tasks, particularly delay minimization in PASS-assisted MEC systems. To further improve performance, we propose a LBPPO algorithm to solve the optimization problem \textbf{P} in the PASS-enhanced MEC framework by leveraging load information of the network.

\begin{figure}[t]
\centering
\includegraphics[scale=0.7]{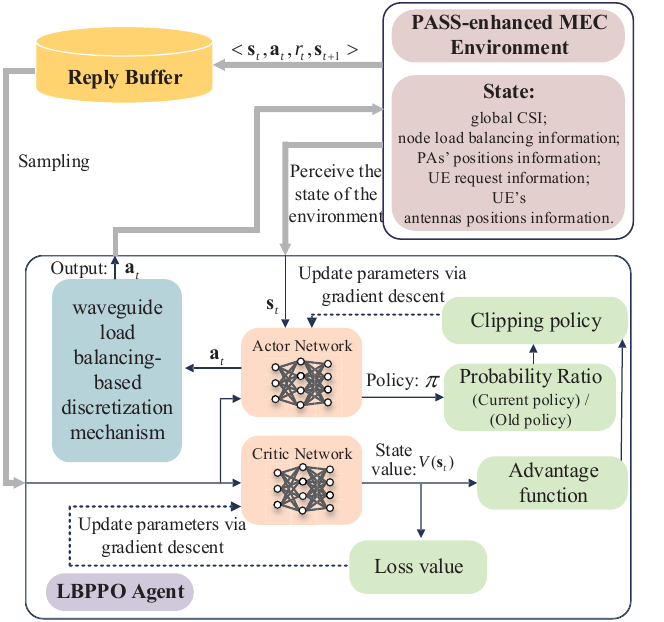}
\caption{Flow diagram of the LBPPO algorithm.}
\label{fig:LBPPO}
\end{figure}

The interaction between the LBPPO agent and the PASS-assisted MEC environment is modeled as a continuous MDP, as shown in Fig.~\ref{fig:LBPPO}. At each time step $t$, the agent observes state $s_t$, samples action $a_t$ from policy $\pi(a_t|s_t)$, transitions to state $s_{t+1}$ according to $P(s_{t+1}|s_t,a_t)$, and receives reward $r_t = R(s_t, a_t)$. The reward function is designed to guide the agent toward minimizing latency under energy constraints. Therefore, the effectiveness of LBPPO in solving problem \textbf{P} critically depends on the design of the state and action spaces, as well as the reward structure. Furthermore, in practical settings where UEs typically outnumber waveguides, multiple users sharing the same waveguide experience increased inter-user interference and degraded SINR. Thus, incorporating load balancing into the state-action design is essential.

\subsubsection{In-Network Node Load Balancing-Aware State Space}

Given that offloading and uplink PASS beamforming are jointly considered in the PASS-enhanced MEC MDP model. Crucially, the state includes real-time in-network node load balancing information (the queue status of all in-network nodes), which plays a vital role in facilitating intelligent offloading and resource allocation decisions. In addition, the state's space includes UE request information, PAs' positions information, UE's antennas positions information, and global CSI, comprising SCI from the UEs to PAs and phase-adjusted duplicates from PAs to theirs associated waveguide's feed point. Thus, the state vector at time slot $t$ is denoted as follow
\begin{align}
\mathbf{s_t} \triangleq \{Q_{k,t}, Q_{b,t}, &L_{k,t}, \mathbf{p}_{m,n,t-1}, \\ \nonumber & \mathbf{p}_{k,l,t}, \mathbf{H}_{k,n,t}, \mathbf{h}_{n,t}\}, \forall m, \forall n, \forall k, \forall l.
\end{align}

It is worth noting that the inherent attributes of the in-network nodes, such as the computational capabilities of the UEs $f_k$ $(\forall k \in \mathcal{K})$ and the BS's MEC server $f_b$, remain constant. Accordingly, we assume that these attributes are always included in the state perceived by the agent.

\subsubsection{In-Network Waveguide Load Balancing-Based Action Space}
The formulation of the action space should consider all the decision variables outlined in the optimization problem \textbf{P}. Specifically, it consists of the UE and waveguide association vector $\alpha$, offloading decision vector $\beta$, and uplink PASS beamforming vector, comprising UE's active beamforming vector $\mathbf{w}$, and the positions of PAs $\mathbf{P}$.

For a single PPO agent, only discrete or continuous variables can be optimized in the decision process. In particular, the association between UEs and waveguides $\alpha$ is inherently a discrete variable, while the rest of the decision variables are continuous variables. To bridge this gap, the conventional discretization strategy divides the entire user waveguide-associated action space equally according to the number of waveguides in PASS. The length of any waveguide $n$ associated with UE $k$ at step $t$ in the conventional associated action space $\Gamma_{k,n,t}^{\text{con}}$ can be expressed as
\begin{align}
\Gamma_{k,n,t}^{\text{con}} = \frac{\varpi_{\text{max}}-\varpi_{\text{min}}}{N}, \forall n, \forall k, \forall t,
\end{align}
where $\varpi_{\text{max}}$ and $\varpi_{\text{min}}$ represent the feasible range for the DRL optimization variables.

In response to the challenge that the optimization problem \textbf{P} is relatively difficult to converge, we propose an in-network waveguide load balancing-based discretization mechanism in early stage of training, where the boundaries of the continuous variables are dynamically adjusted according to the real-time load distribution, thereby enabling balanced user-waveguide associations. The length of association decision variables in the associated action space of in-network waveguide load balancing-based discretization mechanism $\Gamma_{k,n,t}^{\text{LB}}$ can be expressed as
\begin{align}
\Gamma_{k,n,t}^{\text{LB}} = \begin{cases}
\begin{aligned}
&\frac{(\varpi_{\text{max}}-\varpi_{\text{min}})(k-\Phi_{k^\prime,t})(\varepsilon_{\text{train}}-\varepsilon_e)}{N\sum_{k^\prime=1}^k(k-\Phi_{k^\prime,t})\varepsilon_{\text{train}}} \\
&\quad \quad \quad \quad \quad \quad \quad \quad \quad \quad +\frac{\varepsilon_{e}}{\varepsilon_{\text{train}}}\Gamma_{k,n,t}^{\text{con}},
\end{aligned} & \varepsilon_e < \varepsilon_{\text{train}}, \\
\Gamma_{k,n,t}^{\text{con}}, & \varepsilon_e \geq \varepsilon_{\text{train}},
\end{cases}
\end{align}
where $\Phi_{n,t}$ denotes the number of UEs associated with the $n$-th waveguide at the $t$-th step. Moreover, $\varepsilon_{\text{train}}$ and $\varepsilon_e$ are the current episode and the episode numbers in the early stages of training. Employing an in-network waveguide load balancing-based discretization mechanism only during the early training stage enhances the exploration capability of the algorithm while ensuring convergence in the later stage. The probability that $n$-th waveguide is associated with the $k$-th user at the $t$-th step selected is $\Gamma_{k,n,t}^{\text{LB}}/({\varpi_{\text{max}}-\varpi_{\text{min}}})$. The mapping relationship between UE-waveguide association decision vector $\boldsymbol{\alpha}_{t}$ and the association decision vector output by the LBPPO agent is
\begin{align}
\boldsymbol{\alpha_{t}} = \left[\alpha_{1,t}, ..., \alpha_{k,t} \right] \leftarrow \boldsymbol{a_{t}^{\alpha}}=\left[\varpi_{1,t}, ..., \varpi_{k,t} \right],
\end{align}
\begin{align}
\alpha_{k,t} = n \leftarrow \sum_{n^\prime=1}^{n-1}\Gamma_{k,n^\prime,t} \geq \varpi_{k,t} \geq \sum_{n^\prime=1}^{n} \Gamma_{k,n^\prime,t}.
\end{align}

Therefore, all actions in the in-network waveguide load balancing-based action space can be formulated as
\begin{align}
\mathbf{a_t} = \{\mathbf{w_t}, \mathbf{P_t}, \boldsymbol{\alpha_t}, \boldsymbol{\beta_t}\}, \forall t.
\end{align}

\subsubsection{Reward Function}
The primary objective of DRL is to identify an optimal policy that maximizes the cumulative reward received by the agent through sequential state transitions. The design of the reward function plays a critical role in guiding the exploration process for solving Problem \textbf{P} and significantly influences the convergence behavior of the algorithm. To jointly optimize uplink PASS beamforming and intelligent offloading while satisfying user service requirements as specified in constraint (22e), these aspects are incorporated into the reward function. Accordingly, the reward function at step $t$ is formulated as follows
\begin{align}
r_t = \sum_{\substack{k \in \mathcal{K} \\ t_{k,t} \geq t_{\text{QoS}}}} r_{\text{QoS}} - \phi_t \cdot t_{k,t}.
\end{align}
where $\phi_t$ denotes the penalty coefficient associated with latency, and $r_{\text{QoS}}$ represents the reward granted when the QoS requirements of UEs are satisfied.

\subsubsection{Neural Network Architecture and Training Process:}\label{architecture}
In the PASS-enhanced MEC system, the design of neural networks in LBPPO plays a critical role in accurately capturing system dynamics and ensuring efficient policy optimization. LBPPO employs two networks, namely the policy (actor) and the value (critic) networks. The actor network maps the observed system state to either action probabilities (for discrete spaces) or bounded continuous actions (for continuous spaces).


The critic network estimates the state-value function $V(s)$to guide policy updates. The actor network consists of three fully-connected layers with 64 neurons and Tanh activation, where the third layer outputs the mean of the Gaussian action distribution for continuous control. The critic network also has three hidden layers with 64 neurons and ReLU activation. Its input dimension equals the sum of the state and action dimensions, and the output is the corresponding Q-value. The input and output layers of both networks are aligned with the state and action space dimensions of the system. Both networks share a similar structure: an input layer matching the state dimension, two hidden layers with 64, 128, and 64 neurons and ReLU activation, and an output layer tailored to their specific roles. The actor uses a Tanh output for continuous actions, while the critic outputs a scalar value. 
As the training process of LBPPO follows standard PPO, it is omitted here for brevity.



\vspace{-0.2cm}

\section{Numerical Results} \label{LBPPO}
\vspace{-0.2cm}
In this section, we present the simulation results to evaluate the performance of the proposed PASS-enhanced MEC system optimized by the LBPPO algorithm. We consider a squared region of 10 m $\times$ 10 m, where $K=5$ UEs with $L=3$ antennas are uniformly distributed. The PASS is composed of $N=3$ waveguides, each equipped with $M=4$ PAs. The carrier frequency is $f_c = 28$ GHz, the effective refractive index of the waveguide is $n_{e}=1.4$, the speed of light $c=3\text{e}^8$, the minimum pinching distance is set to $\Delta_l = \lambda/2$, the hight of the waveguide $H = 3$ m, the default max transmit power of UE $P^{\text{max}}=20$ dBm and the noise power $\sigma^2=-90$ dBm. In addition, request duration $\dot{\tau} = 2$ s, the task length is $L_{k,\tau}=2\text{e}^7$ bit, and the computation capacity of UE and BS are $f_k = 1\text{e}^9$ r/s and $f_b = 2\text{e}^9$ r/s, respectively.

For the proposed LBPPO algorithm, the default learning rate of all networks are $1\text{e}^{-3}$. All neural network architectures are introduced in Section \ref{architecture}. The maximum episode and step are set to 2000 and 30, respectively. Moreover, a sampling batch size of 512 is extracted from the replay buffer during the training process of the agent. In the reward function, the reward from satisfying UE QoS requirements $r_{\text{QoS}}=1$, and the  penalty coefficient associated with
latency $\phi_t=1$.

\begin{figure}[h]
\centering
\includegraphics[scale=0.38]{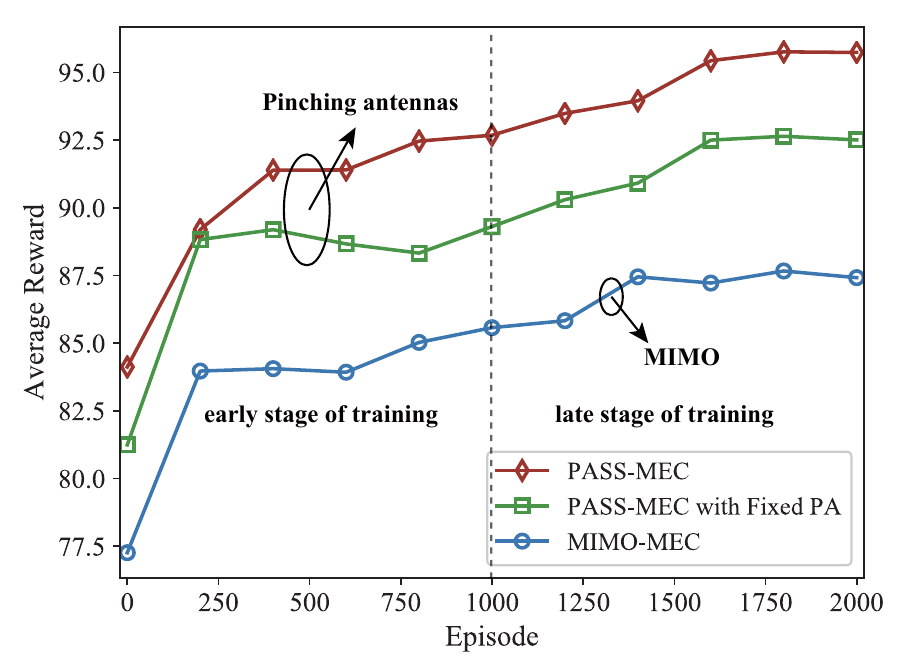}
\vspace{-0.2cm}
\caption{Average reward of different models.}
\label{fig:training}
\vspace{-0.4cm}
\end{figure}

Fig.~\ref{fig:training} presents the average reward obtained by the different models. In the MIMO-assisted MEC model, multiple UEs simultaneously send signals directly to the BS, resulting in high interference. The PASS model, on the other hand, implements SDMA through three waveguides. Since this simulation primarily demonstrates the convergence of the algorithm when solving the model, we increased the reward parameter for the MIMO-assisted MEC system by three times to enhance the visibility of the simulation graph. It can be seen that as the training process progresses, all three models can converge due to the accumulation of experience of the DRL agent. Furthermore, due to the load balancing-based discretization mechanism in LBPPO, the decoding of UE and waveguide association actions is affected by load information in the algorithm's early training stages, resulting in significant fluctuations in the early stages of training. However, this strategy improves the LBPPO algorithm's exploratory capabilities, resulting in the agent having more experience to converge in the later stage of the training.


\begin{figure}[h]
\centering
\includegraphics[scale=0.38]{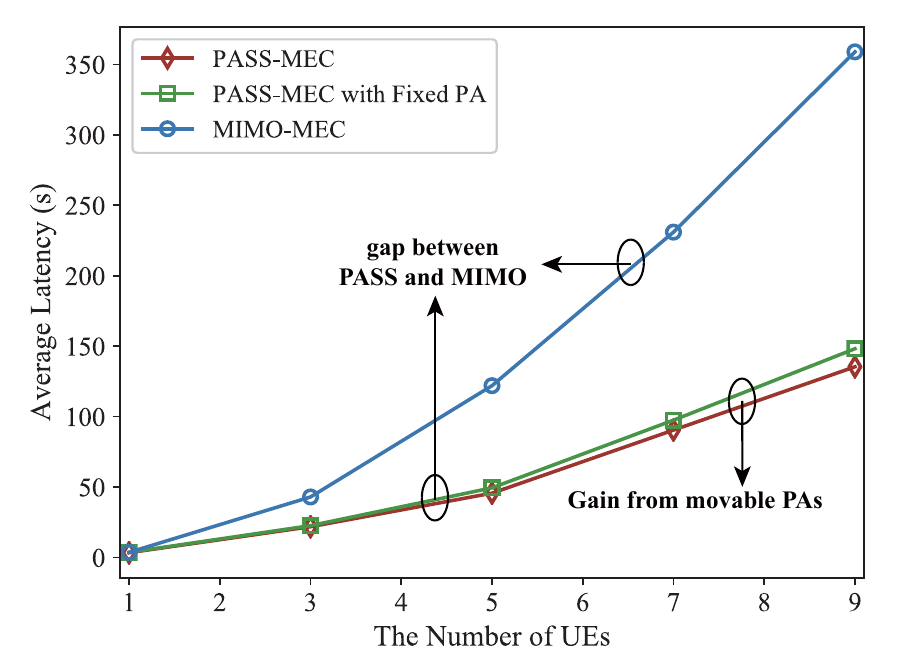}
\vspace{-0.2cm}
\caption{UEs versus average latency.}
\label{fig:UEs}
\vspace{-0.4cm}
\end{figure}

Fig.~\ref{fig:UEs} illustrates the average latency of all schemes versus the number of UEs. It can be observed that PASS-enhanced MEC systems significantly outperforms the conventional MIMO-assisted MEC system. This advantage stems from the inherent SDMA capability of PASS and its ability to replace long-distance propagation with short-distance LoS links. Moreover, compared with the PASS-assisted MEC model with fixed PAs, movable and adjustable PAs can adaptively tune the uplink PASS beamforming according to UE requests, UE locations, and network states, thereby achieving improved performance. 
With more UEs, both inter-user interference and BS queuing delay increase, resulting in higher response latency.

\begin{figure}[h]
\centering
\includegraphics[scale=0.38]{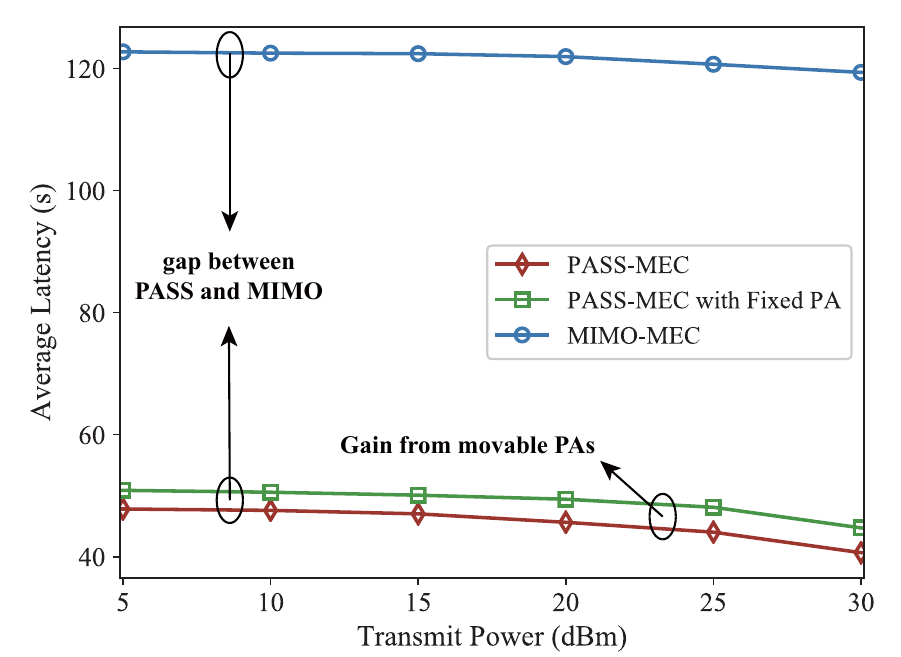}
\vspace{-0.2cm}
\caption{Transmit power versus average latency.}
\label{fig:TP}
\vspace{-0.4cm}
\end{figure}

Fig.~\ref{fig:TP} displays the average latency of all schemes versus different transmit power. As the transmit power of UEs increases, the average response latency of UE requests decreases across all models. Compared with the MIMO-assisted MEC system, the LBPPO agent achieves a greater reduction in average latency by jointly optimizing uplink PASS beamforming and offloading decisions.
\section{Conclusion} \label{conclusion}
In this paper, we proposed a PASS-enhanced MEC system to minimize network latency by jointly optimizing uplink PASS beamforming and task offloading. To address the instability and slow convergence caused by the $\max$ operator in the objective function, we developed a LBPPO algorithm that incorporates both node-level and waveguide-level load balancing information into the policy design. Simulation results demonstrated that PASS-enhanced MEC shows stronger convergence ability than PASS-enhanced MEC with fixed PAs and traditional MIMO-assisted MEC, especially in scenarios with a large number of UEs and high transmit power. These results highlight the potential of combining PASS with DRL-based optimization for future low-latency and high-reliability MEC systems.

\small
\bibliography{mybib}
\bibliographystyle{IEEEtran}

\end{document}